# The Bounded Anharmonic Oscillators: a simple approach

## F. Maiz[1] and M. Nasr[2]

[1]IPEIN, B.P:62, Merazka 8000, Nabeul, Tunisia.
[2]Faculty of mathematical sciences, university of Khartoum, p.o. box: 32, Sudan.
Actual Address:
[1]Department of Physics, College of Sciences, King Khalid University, Saudi Arabia.
Email: fethi_maiz@yahoo.fr; fmaiz@kku.edu.sa
[2]Department of Mathematics, College of Sciences, King Khalid University, Saudi Arabia.
 Email: mednasr@kku.edu.sa; m_nasr@yahoo.com

*Abstract:*

*We have developed a simple method to solve anharmonic oscillators equations. The idea of our method is mainly based on the partitioning of the potential curve into (n+1) small intervals, solving the Schrödinger equation in each subintervals, and writing the continuity conditions of each solutions at each subintervals boundary, which leads to the energy quantification condition, so to the energy levels, and finally, one can calculate the exact wave function associated to each energy level. Our method has been applied to three examples: the well-known square well, the bounded and unbounded harmonic oscillators, the Morse potential, and some anharmonic oscillators bounded by two infinite walls. Our method is more realistic, simpler, with high degree of accuracy, both satisfactory and not computationally complicated, and applicable for any forms of potential. Our results agree very well with the preceding ones.*

*1. Introduction*

*One of the challenging problems in nonrelativistic quantum mechanics is to find exact solutions to the Schrödinger equation for potentials that proves to be useful to model phenomena in nuclear physics, solid state physics, molecular-atomic physics, and laser theory. Apart from this, anharmonic oscillators are themselves interesting since the real world deviates from an idealized picture of harmonic oscillators because of self interactions and interactions between them. Therefore, anharmonic systems have been studied extensively both analytical and numerical. The spectra of many important potential functions frequently encountered in quantum mechanics cannot be obtained exactly. Moreover, in most cases, the conventional approximate methods commonly discussed in most standard textbooks [1], are either unsatisfactory or computationally complicated.*

*Many techniques have been developed for determining the energy spectra of various anharmonic oscillators. Highly accurate approximations have been achieved by use of the Hill determinant [2-3], the Bargmann representation [4-5], the coupled cluster method [6-7], the variational-perturbation expansion [8-9], and many other approaches [10-13]. Alhendi [14] used a power-series expansion to calculate the eigenvalues of anharmonic oscillators bounded by two infinite walls, and showed that for large finite values of the separation of the walls, the calculated eigenvalues are of the same high accuracy as the values recently obtained for unbounded case by the inner-product quantization method. Fernández [15] calculated accurate eigenvalues of a bounded oscillator by means of the Riccati–Padé method that is based on a rational approximation to a regularized logarithmic derivative of the wavefunction. Sequences of*

*roots of Hankel determinants approach the model eigenvalues from below with a remarkable convergence rate.*

*The well-known deep double-well potential is a sufficiently important example in this respect, in which splitting of energy levels of two neighboring states, in particular the ground and first excited state, is extremely small because of the non-degenerate nature of energy eigenvalues of bound states for one-dimensional potentials [1,16]. Perturbation, semiclassical, and variational methods are of limited use in such a case. A need thus arises for a relatively simple and effective approximate method with a high degree of accuracy. A variant of approximate methods and numerical techniques has recently been devised to calculate to high precision the spectrum of the one-dimensional symmetric anharmonic oscillators, with either a symmetric solution [17-21], or a symmetry-broken solution [22-24].*

*In this paper we propose a very simple method which allows the determination of the energy levels for any forms of potential. This method is with high degree of accuracy and is both satisfactory and not computationally complicated. This paper is organized as follows. In Sect. 2, we present our simple approach which is based on the partitioning of the potential curve into n+1 small intervals, solving the Schrödinger equation in each subintervals, and writing the continuity conditions of each solutions at each subintervals boundary, which leads to the energy quantification condition, so to the energy levels, and finally, we study the cases of n=0 and n=1, and the well-known infinite square well to introduce the method. In Sect.3, first, we revise the bounded and unbounded harmonic oscillators; we show that for large finite values of the separation of the walls, the calculated eigenvalues are of the same high accuracy as the values recently obtained for the unbounded case. Second, we study the Morse potential in order to test the*

*convergence of each solution. Finally, we use the same application as Alhendi [14] to calculate the eigenvalues of some anharmonic oscillators bounded by two infinite walls.*

## 2. The simple approach

*Formulation*

*To calculate the eigenvalues of energy of harmonic and anharmonic oscillators bounded by infinity high potentials at* $x = a$ *and* $x = b$, *here* $b - a = 2L$ *the large finite values of the separation of the walls, it is necessary to solve the one-dimensional Schrödinger equation ( throughout this paper, we assume that* $\hbar = 1$ *and* $2m = 1$ *):*

$$\frac{d^2\Psi(x)}{dx^2} + (E - V(x))\Psi(x) = 0 \qquad (1)$$

*Where,* $V(x)$ *is the potential energy.*

*As described by Fig. 1,* $V(x)$ *is a continuous function (solid line) between* $x = a$ *and* $x = b$, *in order to determinate the energy levels, we start by partitioning the interval [a,b] into n+1 small subintervals* $I_i = [x_i, x_{i+1}]$ *each with width* $h$, *where* $h = \frac{(b-a)}{n+1}$ *and* $x_i = a + ih$ *for* $i = 0,...,n$.

*The midpoint of each of these subintervals is given by* $\rho_i = \frac{x_{i+1} - x_i}{2}$. *We evaluate our function* $V(x)$ *at the midpoint of any subinterval, and prolong it by adding two large intervals* $I_{-1}$ *and* $I_{n+1}$ *i.e.:*

$$\begin{cases} \text{for } x \leq a, & x \in I_{-1}: \quad V(x) = \infty \\ \text{for } x_i \leq x \leq x_{i+1}, & x \in I_i: \quad V(x) = V(\rho_i) \\ \text{for } b \leq x, & x \in I_{n+1}: \quad V(x) = \infty \end{cases} \qquad (2)$$

*Let* $k_i^2 = (V_i - E)$, *writing and solving the above equation (1) in each subinterval leads to the following solutions:*

$$\Psi_i(x) = X_i \exp(k_i x) + Y_i \exp(-k_i x) \qquad (3)$$

Where $X_i$ and $Y_i$ are constants. For the two intervals $I_{-1}$ and $I_{n+1}$, the potential is infinity, and $\Psi(a) = \Psi(b) = 0$. The functions $\Psi_i(x)$ are twice derivable with respect to x, and they are continuous. The continuity of the solutions $\Psi_i(x)$ and their derivates at the different points $x_i$ allows to the elimination of $X_i$ and $Y_i$, and leads to the following equation known as the **energy quantification condition:**

$$B_n(E) = a_n P_n + b_n Q_n = 0 \qquad (4)$$

Here:

$$C_{ij} = C_{i,j} = \sqrt{(V_i - E)/(V_j - E)}, \quad a_i = e^{-k_i h}, \quad b_i = e^{k_i h}, \quad Q_0 = -1, \quad P_0 = 1,$$

$$P_i = (c_{i-1,i} + 1) a_{i-1} P_{i-1} + (c_{i-1,i} - 1) b_{i-1} Q_{i-1}, \text{ And } Q_i = (c_{i-1,i} - 1) a_{i-1} P_{i-1} + (c_{i-1,i} + 1) b_{i-1} Q_{i-1}$$

One can proof this expression in the general case by the recurrence method as described by Maiz [25, 26]. The energy levels are obtained by the energy values for which the curve of $B_n$ meets the energy axis. We consider the function F(E) defined as: F(E)=signum( $B_n(E)$ ), where the signum function computes the sign of the leading coefficient of the expression [if $x \neq 0$ then $\text{signum}(x) = \dfrac{x}{\text{abs}(x)}$, and signum(0)=0]. The energy levels are indicated by a vertical segments perpendicular to the energy axis which constitute the curve of the function F(E). In this case, the energy levels are determinated with a great precision.

**The case of n=0: The infinite square well**

In physics, the infinite square well is a problem consisting of a single particle inside of an infinitely deep potential well, from which it cannot escape, and loses no energy when it collides with the walls of the box. The problem becomes very interesting when one attempts a quantum-mechanical solution, since many fundamental quantum mechanical concepts need to be introduced in order to find the solution. The problem may be expressed in any number of dimensions, but the simplest problem is one dimensional, while the most useful solution is a particle in the three dimensional box.

We have considered the case of infinite square well (n=0). The Schrödinger's equation may be described as:
$$\frac{\partial^2 \Psi_0}{\partial x^2} = K_0^2 \, \Psi_0 \quad \text{for } x \in I_0, \text{ where } k_0^2 = (V_0 - E)$$
The acceptable solution of this equation is:

$$\Psi_0 = U e^{K_0 x} + V e^{-K_0 x}$$

Using the boundary condition at each interface such as $\Psi_0(x=a) = 0$ and $\Psi_0(x=b) = 0$ and the elimination of U and V lead to the following equality:

$$B_0(E) = a_0 P_0 + b_0 Q_0 = 0$$

The equality $B_0(E) = 0$ is the energy quantification condition, here:

$$a_0 = e^{-k_0 h}, \quad b_0 = e^{k_0 h}, \quad Q_0 = -1, \text{ and } P_0 = 1.$$

The eigenenergies of a particle in a one dimensional box of length 2L are for $p = 1, 2..$ [27]: $E_p = p^2 \pi^2 / (2L)^2$

We notice that the energy levels increase with the integer number p, and decrease with the separation value of the walls. In Fig 2, the energy levels are determinated graphically by using the variation of the function nF(E) as a function of energy for n=1, n=2, and 2L=10, we have chosen to represent nF(E) instead of F(E) to have a clear figure. Our results agree very well with the theoretical ones. The chosen of the n values is to verify the validity of the repartitioning idea, and their boundary conditions. Also, the value of 2L=10 is to be in the vicinity of the reality of the infinite square well problem.

**The case of n=1:**
We have considered the case (n=1). The Schrödinger's equation may be described as:
$$\frac{\partial^2 \Psi_0}{\partial x^2} = K_0^2 \, \Psi_0 \quad \text{for } x \in I_0, \text{ and } \quad \frac{\partial^2 \Psi_1}{\partial x^2} = K_1^2 \, \Psi_1 \quad \text{for } x \in I_1$$
Where $k_i^2 = (V_i - E)$
The acceptable solution of this equation is:
$$\Psi_0 = U e^{K_0 x} + V e^{-K_0 x}$$

$$\Psi_1 = Ue^{K_1 x} + Ve^{-K_1 x}$$

*Using the boundary conditions at each interface such as,*

$\Psi_0(x = a) = 0$, $\Psi_1(x = b) = 0$, $\Psi_0'(x = (a+h)) = \Psi_1'(x = (a+h))$, *and the elimination of U,*

*V, X, and Y lead to the following equality:*

$$B_1(E) = a_1 P_1 + b_1 Q_1 = 0$$

*The equality* $B_1(E) = 0$ *is the energy quantification condition, here:*

$C_{01} = C_{0,1} = \sqrt{(V_0 - E)/(V_1 - E)}$, $a_i = e^{-k_i h}$, $b_i = e^{k_i h}$, $Q_0 = -1$, $P_0 = 1$,

$P_1 = (c_{0,1} + 1)a_0 P_0 + (c_{0,1} - 1)b_0 Q_0$, *and* $Q_1 = (c_{0,1} - 1)a_0 P_0 + (c_{0,1} + 1)b_0 Q_0$

### *3. Applications and discussions*

#### *3.1. The one-dimensional harmonic oscillator*

*The quantum harmonic oscillator is the quantum mechanical analogue of the classical harmonic oscillator. It is one of the most important model systems in quantum mechanics because an arbitrary potential can be approximated as a harmonic potential at the vicinity of a stable equilibrium point. Furthermore, it is one of the few quantum mechanical systems for which a simple exact solution is known. In the one-dimensional harmonic oscillator problem, a particle is subject to a potential* $V(x) = x^2$. *The eigenenergies of the particle are for* $p = 0,1,2,3....$ *[28]:* $E_p = 2(p + \frac{1}{2})$.

*This energy spectrum is remarkable for three reasons. Firstly, the energies are "quantized", and may only take the discrete values 1, 3, 5, and so forth. This is a feature of many quantum mechanical systems. Secondly, the lowest possible energy is not zero, but 1, which is the ground state energy. The final reason is that the energy levels are equally spaced, unlike the infinite square well case (section 3.1.). Table 1. shows the first energy levels for the bounded harmonic oscillator derived using our method for 2L=2,4,10,20, and n=10,30,200,2000. For 2L=20 and n=2000, we obtained the predicted theoretical*

*values for the unbounded harmonic oscillator (section 3.2.1): 1,3,5,7.. . For fixed value of 2L>2 the energy levels values decrease, when the n values increase, because we approach the actual feature of the potential curve. For fixed value of n and 2L>2 the energy levels values decrease, when the 2L values increase, because we move towards the unbounded potential curve.*

*For 2L=2, the potential curve is more bounded, to a degree that we can not consider it as harmonic oscillator (i.e. the curve is very restricted), which explain the increase of the first energy levels.*

### *3.2. The Morse potential*

*For the case of the well-known Morse potential $V(x) = V_0(1-\exp(-\lambda x))^2$, where $V_0$ and $\lambda$ are the depth and range parameters, respectively, the exact energy eigenvalues are given by:*

$$E_n = 2\lambda\sqrt{V_0}\left[(n+\frac{1}{2}) - (n+\frac{1}{2})^2 \frac{\lambda}{2\sqrt{V_0}}\right].$$

*$E_n(V_0 = 400, \lambda = 1) = 19.75, 57.75, 93.75, 127.75,.....$ These energies levels were obtained exactly by Alhendi [14], for a bounded Morse potential confined to the interval $-2 \leq x \leq 2$.*

*In order to test the convergence of each solution, we have represented the variation of the first energy level for the Morse potential as a function of the integer number n. we note that, for low values of n ($n \leq 50$), $E_0$ values are more than 20. The $E_0$ value decreases with the n value and tend asymptotically to the exact value $E_0 = 19.75$, and the curve is very smooth(see Fig.4).*

### *3.3. The bounded one-dimensional anharmonic oscillator*

*In order to apply our method to bounded anharmonic oscillator, we use the same application as Alhendi [14], for calculating the eigenvalues of anharmonic oscillators bounded by two infinite walls.*

*Table 2. shows the energies of the ground and first excited state for the bounded harmonic oscillator, and quartic, sextic, octic, dectic, and duodectic anharmonic oscillators resulting by our method for 2L=2,4, and n=2000. For 2L=2, generally, we obtained the calculated values by Alhendi [14]; we notice that when the anharmonicity degree increases the values of the energy level decrease, and always greater then those values of the harmonic oscillator. In fact, when we pass from the bounded quartic to the bounded duodectic anharmonic oscillators the restriction of the wells by the infinite walls increases, so the values of the eigenenergy increase, see Fig.3. For 2L=4, we find, for the ground state, the same value that has been calculated by Alhendi [14]; but, for the first excited state, we found the underlined value (<u>4.695</u>); it is different from the one obtained by Alhendi [14], which is (4.58734092). We can't understand why the first excited energy level increases when the finite value of the separation of the walls increases. In fact, Alhendi [14] obtained for the first excited energy level the values 4.58734092 for L=2 and 4.648812 for L=3. We observe that when the anharmonicity degree increases the values of the energy level increase, and always greater then those values of the harmonic oscillator. These values are small than those for 2L=2 because the wells are less bounded.*

*4. Conclusion*

*In this paper, we have presented a simple method to solve anharmonic oscillators' equations. The idea of our method is mainly based on the partitioning of the potential curve into (n+1) small intervals, solving the Schrödinger equation in each subintervals, and writing the continuity conditions of each solutions at each subintervals boundary, which leads to the energy quantification condition, so to the energy levels, and finally, one can calculate the exact wave function associated to each energy level. Our method has been applied to three examples: first for the well-known square well, we recover the theoretical predicted values for the great separation value of the walls, second the study of*

*the bounded and unbounded harmonic oscillators, permit to forecast and calculate the energy values of the predicted first energy levels, and finally, calculate the ground and first excited state for the bounded harmonic oscillator, and quartic, sextic, octic, dectic, and duodectic anharmonic oscillators bounded by two infinite walls. In this case, also, we recuperate the preceding found results. Our method is more realistic, simpler, with high degree of accuracy, both satisfactory and not computationally complicated, and applicable for any forms of potential.*

*Figures:*

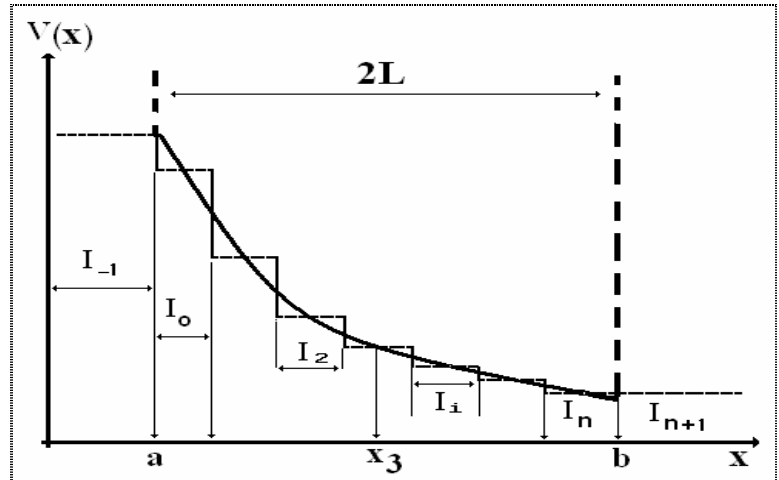

*Fig. 1: repartitioning of the potential curve into (n+1) small intervals.*

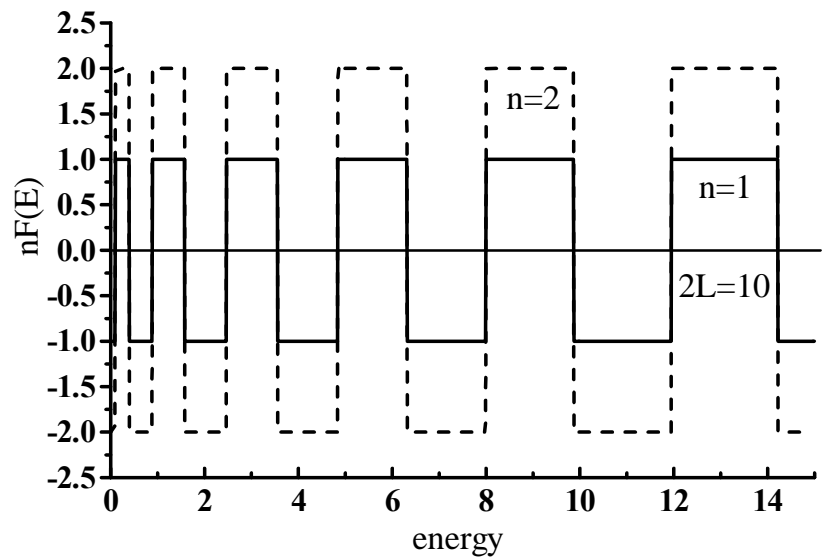

*Fig. 2: nF(E) variations as function of the energy E.*

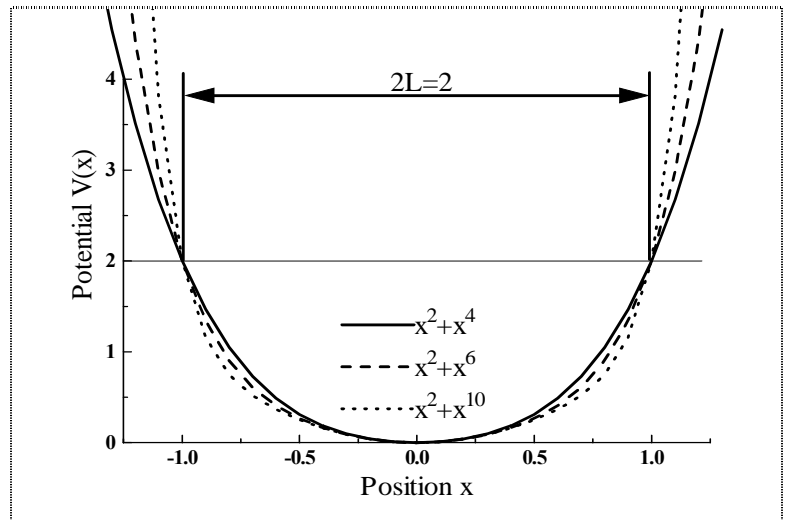

*Fig. 3: variation of the quartic, sextic, dectic anharmonic oscillators potentials as a function of the position x.*

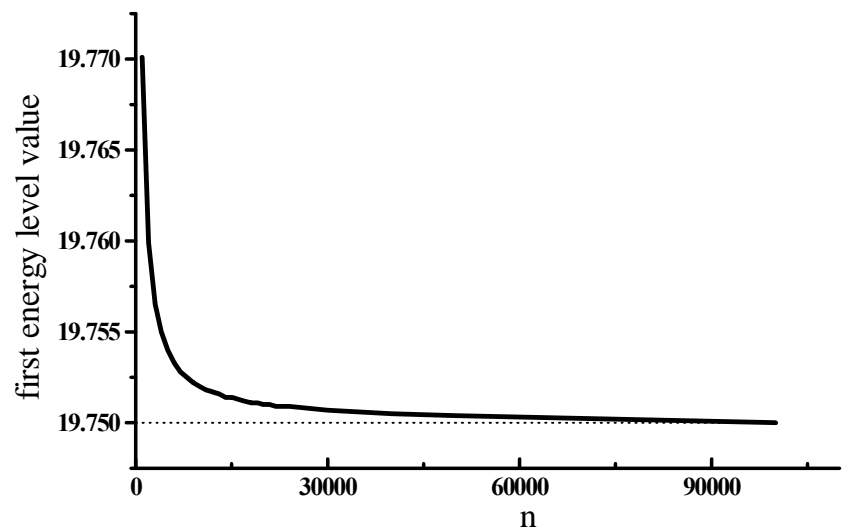

*Fig. 4: variation of the first energy level for the Morse potential as a function of the integer number n.*

*Tables:*

|  | **n=10** | **n=30** | **n=200** | **n=2000** |
|---|---|---|---|---|
| **energy levels (L=1)** | 2.62625<br>10.21325 | 2.59925<br>10.17275 | 2.59925<br>10.15925 | 2.59718<br>10.151348 |
| **energy levels (L=2)** | 1.17551<br>3.749508<br>7.077505<br>11.458502 | 1.09751<br>3.593508<br>6.882505<br>11.263502 | 1.08451<br>3.541508<br>6.817506<br>11.185502 | 1.07151<br>3.52626<br>6.79806<br>11.171385 |
| **energy levels (L=5)** | 1.18175<br>3.36875<br>5.55575<br>7.79675<br>9.86225<br>12.29225 | 1.04675<br>3.11225<br>5.17775<br>7.24325<br>9.30875<br>11.37425<br>13.43975 | 1.00625<br>3.01775<br>5.02925<br>7.04075<br>9.03875<br>11.05025<br>13.06175 | 1.000586<br>3.001496<br>5.002488<br>7.003419<br>9.004527<br>11.005669<br>13.007708 |
| **energy levels (L=10)** | 1.19525<br>3.97625<br>5.73125<br>7.00025<br>9.71375<br>12.34625 | 1.07375<br>3.13925<br>5.20475<br>7.27025<br>9.33575<br>11.40125<br>13.46675 | 1.00625<br>3.01775<br>5.02925<br>7.04075<br>9.05225<br>11.05025<br>13.06175 | 1.000586<br>3.001487<br>5.002569<br>7.003471<br>9.004435<br>11.005464<br>13.007584 |

*Tab. 1: The first energy levels for the bounded harmonic oscillator derived using our method for 2L=2,4,10,20, and n=10,30,200,2000.*

| **n=2000** | $V(x) = x^2 + x^4$ | $V(x) = x^2 + x^6$ | $V(x) = x^2 + x^8$ | $V(x) = x^2 + x^{10}$ | $V(x) = x^2 + x^{12}$ | $V(x) = x^2$ |
|---|---|---|---|---|---|---|
| **L=1** | 2.635<br>10.265 | 2.615<br>10.205 | 2.605<br>10.185 | 2.605<br>10.175 | 2.605<br>10.165 | 2.597<br>10.151 |
| **L=2** | 1.395<br><u>4.695</u> | 1.435<br>5.035 | 1.495<br>5.375 | 1.545<br>5.665 | 1.595<br>5.915 | 1.071<br>3.526 |

*Tab. 2: The energies of the ground and first excited state for the bounded harmonic oscillator, and quartic, sextic, octic, dectic, and duodectic anharmonic oscillators resulting by our method for 2L=2,4, and n=2000.*